\documentclass{article}

\PassOptionsToPackage{authoryear, sort}{natbib}

 \usepackage[preprint]{neurips_2024}
\newcommand{\method}{DyG-RAG}
\usepackage{algorithm}
\usepackage{natbib}
\usepackage{marvosym}
\usepackage{amsthm}
\usepackage{pifont}
\usepackage{enumitem}
\usepackage{amsmath}
\usepackage{amssymb}
\usepackage{color}
\usepackage{booktabs}
\usepackage{multirow}
\usepackage{makecell}
\usepackage{colortbl}

\usepackage{bbding}
\usepackage{adjustbox}
\usepackage{tikz}
\usepackage{subfigure}
\usepackage{enumitem}
\usepackage{url}
\usepackage{caption}  

\definecolor{darksalmon}{rgb}{0.91, 0.59, 0.48}
\definecolor{emerald}{rgb}{0.31, 0.78, 0.47}
\definecolor{pigment}{rgb}{0.0, 0.65, 0.31}
\definecolor{amaranth}{rgb}{0.9, 0.17, 0.31}
\definecolor{iris}{rgb}{0.35, 0.31, 0.81}
\definecolor{uu}{rgb}{0.95, 0.51, 0.51}
\definecolor{spirodiscoball}{rgb}{0.06, 0.75, 0.99}
\definecolor{orange}{rgb}{0.92,0.38,0.25}
\definecolor{crimson}{rgb}{0.86, 0.08, 0.235}
\definecolor{darkblue}{rgb}{0.184, 0.33, 0.59}

\newtheorem{myDef}{Definition}

\usepackage{newfloat}
\usepackage{listings}

\usepackage{import}%
\usepackage{graphicx}%
\usepackage{comment}%
\usepackage{multirow}%
\usepackage{amsmath,amssymb,amsfonts}%
\usepackage{amsthm}%
\usepackage{mathrsfs}%
\usepackage[title]{appendix}%
\usepackage{xcolor}%
\usepackage{textcomp}%
\usepackage{manyfoot}%
\usepackage{booktabs}%
\usepackage{array}
\usepackage{algorithm}%
\usepackage{algorithmicx}%
\usepackage{algpseudocode}%
\usepackage{listings}%
\usepackage{longtable}%
\usepackage[most]{tcolorbox}%
\usepackage{booktabs}
\usepackage{enumerate}
\usepackage{url}

\usepackage{hyperref}
\usepackage{tikz}
\usepackage{pgfplots}
\usepackage{array} 
\usetikzlibrary{shapes, arrows, positioning, fit}
\usepackage{tabularx}
\usepackage{placeins}

\hypersetup{
    colorlinks=true,
    allcolors=[rgb]{0 0 0.5}
    }

\title{DyG-RAG: Dynamic Graph Retrieval-Augmented Generation with Event-Centric Reasoning}

\author{%
    Qingyun Sun\textsuperscript{1}\quad Jiaqi Yuan\textsuperscript{1}\quad Shan He\textsuperscript{1}\quad Xiao Guan\textsuperscript{1}\quad Haonan Yuan\textsuperscript{1}\\
    \textbf{Xingcheng Fu\textsuperscript{2}}\quad \textbf{Jianxin Li\textsuperscript{1}}\textsuperscript{\Letter}
    \quad \textbf{Philip S. Yu\textsuperscript{3}}
    \\
  \textsuperscript{1}Beihang University~~
  \textsuperscript{2}Guangxi Normal University~~
  \textsuperscript{3}University of Illinois Chicago\\
  \small\texttt{\{sunqy,yuanjq,heshan25,xiaoguan,yuanhn,lijx\}@buaa.edu.cn, } \\ \texttt{fuxc@gxnu.edu.cn, psyu@uic.edu}
}
\begin{document}

\maketitle

\begin{abstract}
Graph Retrieval-Augmented Generation has emerged as a powerful paradigm for grounding large language models with external structured knowledge. 
However, existing Graph RAG methods struggle with temporal reasoning, due to their inability to model the evolving structure and order of real-world events. 
In this work, we introduce \method, a novel event-centric dynamic graph retrieval-augmented generation framework designed to capture and reason over temporal knowledge embedded in unstructured text. 
To eliminate temporal ambiguity in traditional retrieval units, \method~proposes Dynamic Event Units (DEUs) that explicitly encode both semantic content and precise temporal anchors, enabling accurate and interpretable time-aware retrieval.
To capture temporal and causal dependencies across events, \method~constructs an event graph by linking DEUs that share entities and occur close in time, supporting efficient and meaningful multi-hop reasoning. 
To ensure temporally consistent generation, \method~introduces an event timeline retrieval pipeline that retrieves event sequences via time-aware traversal, and proposes a Time Chain-of-Thought strategy for temporally grounded answer generation.
This unified pipeline enables \method~to retrieve coherent, temporally ordered event sequences and to answer complex, time-sensitive queries that standard RAG systems cannot resolve. 
Extensive experiments on temporal QA benchmarks demonstrate that \method~significantly improves the accuracy and recall of three typical types of temporal reasoning questions, paving the way for more faithful and temporal-aware generation. 
\method~is available at \url{https://github.com/RingBDStack/DyG-RAG}.
\end{abstract}
\section{Introduction}
\label{sec:intro}
Recent advances in Large Language Models (LLMs) have demonstrated impressive capabilities in many applications~\citep{yao2024survey}. 
However, despite their generalization ability, LLMs remain limited by their parametric knowledge, often struggling with factual consistency and adaptability to evolving information. 
Retrieval-Augmented Generation (RAG)~\citep{ragsurvey1,ragsurvey2} has emerged as a promising paradigm to mitigate these limitations by incorporating external knowledge through retrieval mechanisms. 
Existing RAG approaches typically rely on static retrieval methods that treat documents as isolated and unstructured items, failing to capture the latent relational structure among retrieved knowledge. 
To address the limitations of treating retrieved documents as unstructured text, Graph Retrieval-Augmented Generation (Graph RAG) methods~\citep{graphragsurvey1} have been proposed. 
These approaches represent knowledge as a structured graph, where nodes correspond to information units (e.g., entities, sentences, or passages) and edges encode explicit relationships such as co-reference, semantic similarity, or knowledge base links. 
By enabling multi-hop traversal and relational reasoning, Graph RAG models enhance the LLMs’ ability to answer complex queries that require structured inference.

However, both RAG and Graph RAG methods share a fundamental limitation: they struggle with modeling the temporal dynamics of knowledge. 
In real-world scenarios, information is often not static but unfolds over time. 
Understanding such temporal evolution is crucial for tasks like timeline generation~\citep{sojitra2024timeline} and historical question answering~\citep{maged2024historyquest}. 
Existing retrieval-augmented methods typically treat knowledge as a static snapshot. 
Vanilla RAG, for instance, relies heavily on semantic similarity between the query and candidate documents, often incorporating timestamps only as an auxiliary dimension in the vector space. 
Even Graph RAG models, while introducing structural representations, fall short in capturing temporal dynamics. 
These approaches lack an explicit representation of temporal ordering or causal progression, making it ill-suited for queries that depend on the relative sequence of events.
In practice, this leads to several critical failure modes:
\ding{182}
\textbf{Semantic retrievers struggle to distinguish temporal differences.} For example, the questions \textit{``What did Barack Obama do \underline{before} 2008?''} and \textit{``What did Barack Obama do \underline{after} 2008?''} yield nearly identical embeddings, as semantic similarity dominates, while temporal intent is ignored.
\ding{183}
\textbf{Temporal constraints are inherently difficult to express and enforce during the retrieval process.} 
Queries like \textit{``What happened after Obama became president?”} require conditioning on a temporal anchor, but standard retrievers lack mechanisms for relative or conditional retrieval, often returning temporally mismatched results.
\ding{184}
\textbf{Lack of temporal compositionality.} 
Questions such as \textit{``How did Obama’s foreign policy evolve after his first term?”} require reasoning over a sequence of events, 
but standard retrievers return isolated chunks without modeling event progression or temporal dependencies, making it difficult to synthesize coherent answers grounded in temporal chains.
A natural extension is to incorporate temporal knowledge graphs (TKGs), where edges are annotated with timestamps to encode when a relation holds. 
However, these graphs are fundamentally limited in their expressiveness for temporal reasoning: timestamped edges capture only the duration of relations, not the evolution of entity states or event sequences.
For example, a TKG may represent that \textit{``Barack Obama–PresidentOf–USA”} holds from 2009 to 2017, but it cannot directly model how his roles, actions, or decisions changed over time, or how they relate to prior and subsequent events.



\begin{figure}[t]
\centering

\subfigure[Comparison of RAG pipeline]{
    \begin{minipage}[t]{0.63\linewidth}
    \centering
    \includegraphics[width=\linewidth]{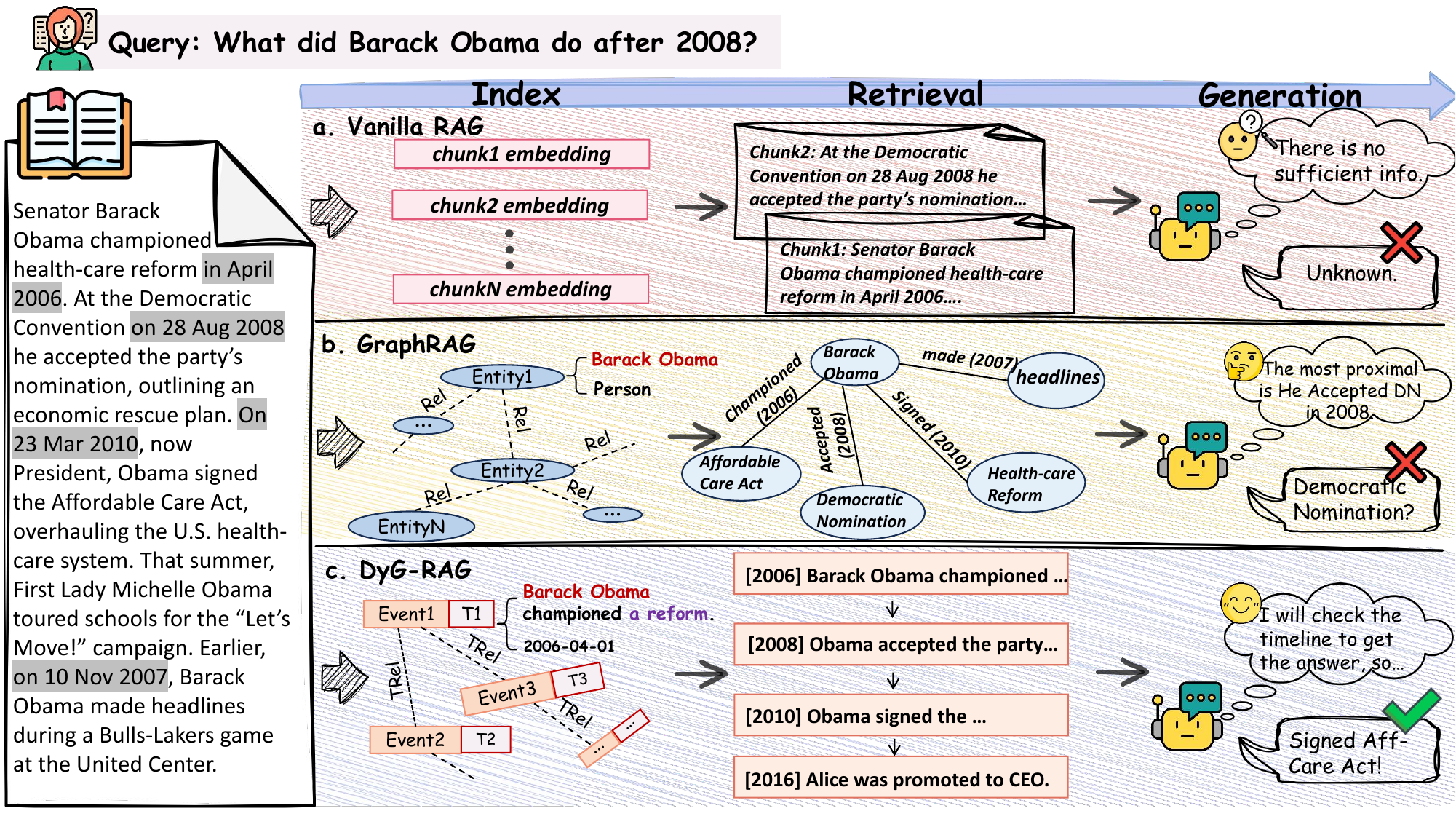}
    \label{fig:pipe_comparison}
    \end{minipage}
}
\hfill
\subfigure[Comparison of multi-dimensions]{
    \begin{minipage}[t]{0.33\linewidth}
    \centering
    \includegraphics[width=\linewidth]{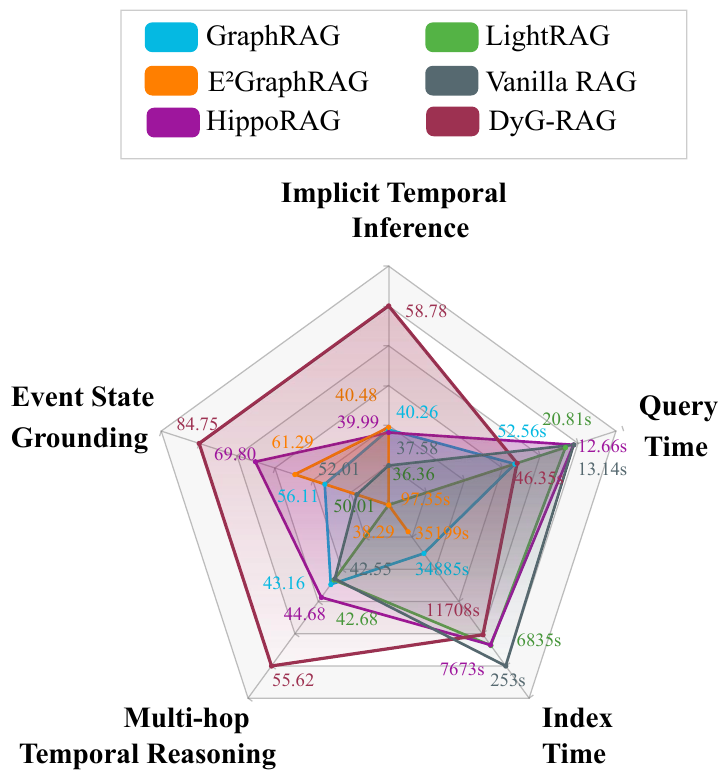}
    \label{fig:radar}
    \end{minipage}
}

\vspace{-0.3cm}
\caption{
Comparison and evaluation of representative RAG methods and \method. 
In Figure~\ref{fig:radar}, three axes including Event State Grounding, Implicit Temporal Inference, and Multi-hop Temporal Reasoning are accuracy metrics (\%), while Query Time and Index Time represent efficiency metrics. 
As illustrated, \method~exhibits superior performance with comparable efficiency, demonstrating a clear advantage over prior RAG methods.
}
\label{fig:comparison}
\end{figure}

In this paper, we present \textbf{\method}, a \textbf{Dy}namic \textbf{G}raph \textbf{R}etrieval-\textbf{A}ugmented \textbf{G}eneration framework that models the knowledge as a dynamic graph from an event-centric perspective. 
To organize temporally grounded knowledge, \method~proposes the Dynamic Event Unit (DEU), which restructures raw text into atomic temporally explicit knowledge units in the event granularity, ensuring that subsequent retrieval is temporally aligned and semantically focused.
To capture temporal and causal dependencies across events, we construct an event graph by linking DEUs that share entities and occur close in time, enabling efficient and meaningful multi-hop reasoning.
To ensure temporally consistent generation, we introduce an event timeline retrieval pipeline that retrieves event sequences via time-aware vector search and traversal, and guides LLMs using a Time Chain-of-Thought strategy for temporally grounded answers. 
The contributions are summarized as follows:
\begin{itemize}[leftmargin=1.5em]
 \setlength{\itemsep}{1pt}
  \setlength{\parskip}{1pt}
  \setlength{\parsep}{1pt}
    \item We propose a novel framework \method, the first dynamic graph retrieval-augmented generation framework that structures and reasons over knowledge from an event-centric perspective. 
    \item By explicitly modeling temporal knowledge as an event graph,
    \method~enables time-sensitive retrieval via event timeline reconstruction and enhances reasoning via a Time Chain-of-Thought prompting mechanism, resulting in more faithful and grounded generation.
    \item Extensive experimental results demonstrate that \method~significantly improves the performance of three typical types of temporal QA tasks. 
\end{itemize}

\section{Related Work}
\label{sec:relatedwork}
\begin{table}[t]
\caption{Representative GraphRAG Methods Comparison}
\label{tab:methods}
\renewcommand\arraystretch{1.2} 
\centering
\resizebox{\linewidth}{!}{
\begin{tabular}{l|l|l|l|l|c}  
\toprule
\textbf{Method} & \textbf{Graph Unit} & \textbf{Edge Type} &
\textbf{Retrieval Strategy} & \textbf{Reasoning Mechanism} &
\textbf{Dynamic} \\
\midrule                             
GraphRAG      & Entity + Community    & KG Relations + Community Links       & Local + Global        & Community summary & \textcolor{orange}{\XSolidBrush}   \\
LightRAG      & Chunk Entities        & Intra-chunk KG Relations        & Dual-level keywords   & Shallow path merge & \textcolor{orange}{\XSolidBrush}   \\
E$^2$GraphRAG & Summary Tree + Entity & Semantic + Hierarchical Links        & 
Adaptive Local/Global & Chunk ranking      & \textcolor{orange}{\XSolidBrush}   \\
HippoRAG      & Concept Nodes         & Concept Associations       &
PPR-guided multi-hop  & PPR subgraph rank  & \textcolor{orange}{\XSolidBrush}   \\
HybridRAG     & KG + Chunks           & KG relations       &
Hybrid merge          & Evidence voting    & \textcolor{orange}{\XSolidBrush}   \\
\midrule                             
\textbf{DyG-RAG} &
\textbf{Dynamic Event Units} &
\textbf{Temporal--Semantic Links} &
\textbf{Time-aware graph walk} &
\textbf{Time-CoT} &
\textcolor{pigment}{\Checkmark} \\
\bottomrule
\end{tabular}}
\end{table}

\subsection{Retrieval-Augmented Generation}
Retrieval-Augmented Generation (RAG) aims to enhance LLMs by combining relevant information retrieved from external sources, enabling factual grounding and reducing hallucinations~\citep{ragsurvey1,ragsurvey2}. 
Subsequent works have extended RAG in various directions, including retriever design~\citep{G-retriever}, retrieval strategies~\citep{cuconasu2024power}, and generation techniques~\cite{finardi2024chronicles}. 
How to enable RAG to retrieve task-relevant and domain-specific knowledge and to perform better reasoning has become a key focus of current research. 
For example, RAG-end2end~\citep{siriwardhana2023improving} extends the vanilla RAG by updating all components with domain-specific knowledge during training. 
Self-RAG~\citep{selfrag} retrieves and reflects using reflection tokens, enabling the generation to tailor to diverse task requirements. 
Despite these advances, existing RAG methods employ static text chunks and focus on semantic relevance, often overlooking temporal structure and event-level reasoning, which our \method~framework aims to address.

\subsection{Graph Retrieval-Augmented Generation}
Graph Retrieval-Augmented Generation (Graph RAG) shows superior performance in reasoning by representing knowledge in a more structural way~\citep{graphragsurvey1,ragvsgraphrag}.
GraphRAG~\citep{graphrag} builds a graph index by deriving an entity knowledge graph from the source documents and uses community summaries to generate the partial responses. 
LightRAG~\citep{lightrag} proposes a dual-level retrieval mechanism from both low-level and high-level knowledge discovery to improve response times.
E$^2$GraphRAG~\citep{E2GraphRAG} constructs a summary tree and an entity graph for knowledge modeling and then constructs bidirectional indexes and an adaptive retrieval strategy to capture the many-to-many relationships efficiently.
HippoRAG~\citep{hipporag} is inspired by human long-term memory and uses personalized PageRank to identify relevant subgraphs for multi-hop reasoning. 
HybridRAG~\citep{hybridrag} combines Graph RAG and vanilla Vector RAG techniques to enhance question-answer systems to generate more accurate and contextually relevant answers.
Some works focus on exploiting the domain-specific knowledge to enhance the LLMs.
However, current Graph RAG approaches typically rely on static or schema-constrained knowledge structures and overlook the evolving nature of real-world information.
This makes it difficult to capture temporally sensitive knowledge or reason over event dynamics, which our proposed \method~aims to address by modeling knowledge as dynamic event graphs.
The summary of representative RAG methods and Graph RAG methods is shown in Table~\ref{tab:methods}.
\subsection{Enhancing LLMs with Temporal Knowledge}
Prior works have shown that LLMs often possess outdated or temporally inconsistent knowledge, which limits their ability to reason about time-sensitive or evolving facts~\citep{piryani2025s}.
To address this, recent efforts have explored augmenting LLMs with temporal knowledge.
A common line of work uses explicit time representations in the generation process of LLMs.
Chain-of-Timeline~\citep{Chain-of-Timeline} introduces topic-relevant event timelines in SQL-style formats to align model attention with time identifiers. 
Recently, another popular line of work incorporates the pre-defined Temporal Knowledge Graphs (TKGs) to enhance LLMs’ temporal reasoning ability. 
These methods typically enhance LLMs via two mechanisms: (1) integrating TKG facts into prompts~\citep{TimeR4,yuan2024back} or (2) aligning language model representations with temporal knowledge graph embeddings~\citep{GenTKGQA}.
TimeR$^4$~\citep{TimeR4} exploits the knowledge in the TKGs in a ``Retrieve-Rewrite-Retrieve-Rerank'' way. 
GenTKGQA~\citep{GenTKGQA} retrieves subgraphs from TKGs and uses the subgraph embeddings to enhance answers. 
ARI~\citep{ARI} extracts relevant information and provides it to LLMs via abstract reasoning induction.

\section{\method}
\label{sec:method}
\begin{figure*}[!tp]
  \centering
  \input{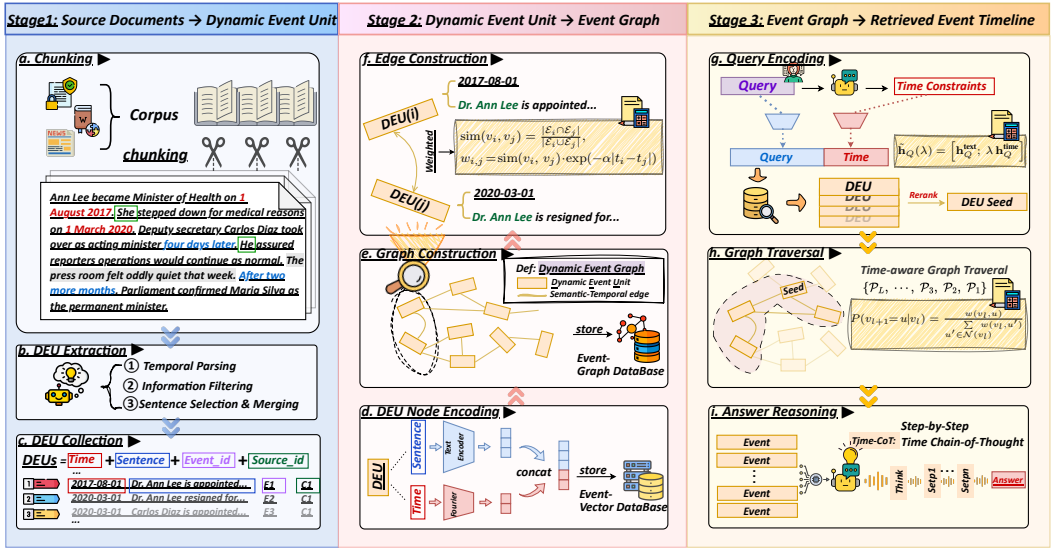}
  \caption{Overall framework of \method. The framework consists of three stages: (1) Source documents are parsed into structured Dynamic Event Units (DEUs) via an LLM; (2) A dynamic event graph is constructed from the DEUs and stored in both vector and graph databases; (3) Given a query, the system performs bi-encoding, DEU seed retrieval, and time-aware graph traversal to generate temporally coherent timeline events and finally reason the answer.
  }
\label{fig:framework}
\end{figure*}

In this section, we systematically introduce the \method~framework, a novel solution that enhances the Graph RAG with dynamic event extraction and reasoning.
The overall framework of \method~is shown in Figure~\ref{fig:framework}. 

\subsection{Preliminary}
\label{sec:preliminary}


Following the setting of general Graph RAG~\citep{graphragsurvey2}, the steps are as follows. 
First, given a text data source $D=\{d_1,d_2,\cdots,d_n\}$, Graph RAG transforms the text data into a graph-structured data source $G$ by a constructor.
Second, the users' query $Q$ is processed by a query processor
and passed to the retriever
to obtain the relevant content $C$. 
Finally, the LLMs generate the final answer using the retrieved content $C'$ as triggers.

\subsection{\method~Workflow}
As an overview, the workflow of \method~converts unstructured text into a dynamic graph by extracting event units, building an interconnected index, and retrieving relevant event sequences. Each stage transitions naturally into the next: from detecting events to indexing relationships to resolving queries, ensuring that the evolving dynamics of knowledge are consistently captured and preserved throughout the entire process.

\textbf{Stage 1: Source Documents $\rightarrow$ Dynamic Event Unit (Section~\ref{sec:eventunit}). }
DyG-RAG decomposes raw text in the source documents $D=\{d_1,d_2,\cdots,d_n\}$ into precisely anchored Dynamic Event Units $\{\rm DEU\}$ that capture essential semantic and temporal information, forming the foundational storage schema for subsequent processing.


\textbf{Stage 2: Dynamic Event Unit $\rightarrow$ Event Graph (Section~\ref{sec:index}). }
DyG-RAG organizes extracted dynamic event units into a knowledge-enriched dynamic event graph $G$ where edges encode both entity co-occurrence and temporal proximity, creating a navigable structure that models narrative flows and causal dependencies.


\textbf{Stage 3: Event Graph $\rightarrow$ Retrieved Event Timeline (Section~\ref{sec:retrieval}).}
DyG-RAG executes a four-phase pipeline: coarse retrieval with time-enhanced embeddings, cross-encoder–based semantic filtering, multi-seed graph traversal, and chronological timeline construction, to reconstruct coherent event sequences and resolve complex event-centric queries.


\subsection{Dynamic Event Unit Extraction}
\label{sec:eventunit}
Existing RAG systems typically index paragraph-level chunks, which often blur temporal boundaries and embed multiple time points, leading to imprecise temporal matching and reduced interpretability. 
To enable temporally grounded retrieval, we replace traditional retrieval units (\textit{e.g.}, paragraphs or knowledge graph triples) with \textbf{Dynamic Event Units (DEUs)}
defined as follows:
\begin{myDef}[Dynamic Event Unit]
A Dynamic Event Unit (DEU) is a self-contained factual statement
that describes a discrete event or stable state occurring at a specific time point or over a clearly defined interval.
\begin{equation}
    {\rm DEU}=\{s_i,t_i,{\rm ID}_{event},{\rm ID}_{source}\},
\end{equation}
where $s_i$ is the sentence,$t_i$ is the normalized timestamp, ${\rm ID}_{event}$ and ${\rm ID}_{source}$ are the event ID and source ID. 
\end{myDef}
This formulation aligns directly with temporal questions (\textit{e.g.}, ``\textit{What happened?}'', ``\textit{When did something change?}'') and serves as the minimal and coherent unit for time-aware retrieval.

The extraction pipeline of DEU consists of four key components: document chunking, temporal parsing, information filtering, and sentence selection and merging. 

\textbf{(1) Document Chunking.}
We first split each source document $d_i \in D$ into overlapping segments of fixed length to preserve contextual coherence and bound computational cost. To mitigate topic drift and maintain semantic consistency, we prepend the document title to each chunk. Documents shorter than the predefined chunk length are retained as single segments. Subsequently, we extract candidate events from each chunk using an LLM for downstream processing.

\textbf{(2) Temporal Parsing.}
We then identify temporal expressions within each candidate event and normalize them to create a consistent time anchor.  
Absolute timestamps (\textit{e.g.}, ``\textit{March 2008}'', ``\textit{2021-06-15}'') are identified and prioritized by granularity. 
The finest-grained date is assigned as $t_i$ and stored in a time stack for anchoring future events.
Relative or vague expressions (\textit{e.g.}, ``\textit{earlier that year}'', ``\textit{recently}'') are resolved by referencing the most recent absolute date within the same paragraph or context window.
Time intervals (\textit{e.g.}, ``\textit{from 2010 to 2015}'') retain their full span in text, but use the earliest point as $t_i$ for indexing.
If no reliable temporal anchor can be extracted, $t_i$ is assigned a static value, indicating timeless background facts.

\textbf{(3) Information Filtering.}
To ensure retrieval-relevant content, we compute an information score for each candidate event sentence based on the presence of key attributes: \ding{172} Contains named entities or coreferable referents. \ding{173} describes a state change or eventive predicate (\textit{e.g}., ``\textit{became}'', ``\textit{resigned}'', ``\textit{launched}''). \ding{174} includes results or quantitative indicators. \ding{175} is anchored in time with month-level precision or higher.
We assign one point for each criterion that the candidate sentence satisfies. 
Only candidates with $score(s)\geq 1$ are retained as DEUs. This step removes generic, underspecified, or off-topic sentences.

\textbf{(4) Sentence Selection and Merging.}
Finally, we normalize, disambiguate, and aggregate valid DEUs to ensure appropriate granularity and semantic coherence.
For coreference resolution, ambiguous pronouns are replaced with explicit entity mentions. 
Regarding event granularity, we assign one DEU per sentence, unless multiple tightly-related actions share the same time anchor, in which case their predicates are merged into a coordinated DEU.
For the temporal resolution constraint, events with different day-level timestamps are not merged to preserve temporal precision.

This process yields a clean, structured set of DEUs, each representing a unique, time-localized event ready for graph-based indexing and retrieval. By matching the natural unit of human temporal questions (\textit{e.g.}, \textit{``what happened?''}, \textit{``when did something change?''}), DEUs serve as the fundamental building block for downstream dynamic graph construction and temporal reasoning.


\subsection{Event Graph Construction and Indexing}
\label{sec:index}
To support structured retrieval and temporal reasoning, we organize the extracted DEUs into a dynamic Event Graph.

\begin{myDef}[Event Graph]
An Event Graph $G=(\mathcal{V},\mathcal{E})$ is weighted graph, where nodes $v_i \in\mathcal{V}$ represent {\rm DEUs} and edges $e_{ij}\in \mathcal{E}$ encode the degree of temporal and semantic relevance between events ${\rm DEU}_i$ and ${\rm DEU}_j$ with edge weight $w_{i,j}\in [0,1]$.
\end{myDef}

\textbf{(1) DEU Node Encoding.} 
Each DEU is encoded into a dense vector representation that fuses semantic and temporal information. Specifically, given a DEU $v_i$ with sentence text $s_i$ and timestamp $t_i$, we compute its embedding as:
\begin{equation}
    \mathbf{z}_i=Concat(\mathbf{h}_i^{text},\mathbf{h}_i^{time}),~\mathbf{h}_i^{text}={\rm Encoder}_{text}(s_i),~\mathbf{h}_i^{time}=\phi(t_i)
\end{equation}
Where $Concat(\cdot,\cdot)$ is the concatenation operation, $\mathbf{h}_i^{text}$ is the sentence embedding generated by a pretrained encoder, and $\mathbf{h}_i^{time}$ is the time embedding. $\mathbf{h}_i^{time}$ is obtained via a Fourier time encoder $\phi(\cdot)$ that maps the timestamp $t_i$ into a smooth periodic representation capturing relative time distances.
    
\textbf{(2) Edge Construction and Weighting.}
The edges between DEU nodes in the event graph are constructed based on two core criteria: \textit{Entity Co-occurrence} and \textit{Temporal Proximity}.
For Entity Co-occurrence, two DEUs must mention at least one common named entity or co-referent entity.
For Temporal Proximity, the absolute time difference between their timestamps must be within a threshold window $\Delta t$.
Formally, an undirected edge $e_{i,j}$ is added between nodes $v_i$ and $v_j$ only if:
\begin{equation}
{\rm EntityOverlap}(e_{i,j})>0~and~|t_i-t_j|\leq\Delta t.
\end{equation}
To capture semantic closeness between two events, we leverage their involved entities for computing the overlap of their respective entity sets:
\begin{equation}
sim\bigr(v_i,\,v_j\bigr)=
\frac{\lvert \mathcal{E}_i \cap \mathcal{E}_j \rvert}
     {\lvert \mathcal{E}_i \cup \mathcal{E}_j \rvert},
\label{eq:sim}
\end{equation}
where $\mathcal{E}_i$ and $\mathcal{E}_j$ denote the sets of named entities extracted from events $v_i$ and $v_j$. 

This entity-based similarity ensures that multi-hop traversals remain consistently anchored around common entities, thereby reducing semantic drift in longer reasoning paths. Moreover, successive overlaps among entities across events naturally reveal meaningful event chains (e.g., \textit{Person} $\rightarrow$ \textit{Organization} $\rightarrow$ \textit{Policy}), effectively enhancing the ability to uncover complex, indirect relationships.

To further incorporate temporal information into the edge weighting, each edge is assigned a weight $w_{i,j}$ that combines semantic similarity and temporal closeness:
\begin{equation}
w_{i,j}=sim(v_i,\,v_j)\cdot\exp(-\alpha|t_i-t_j|),
\label{edge-weight}
\end{equation}
where $\alpha$ is a decay hyperparameter controlling the sensitivity of edge weights to temporal proximity.

To ensure sparsity and focus traversal on meaningful connections, we restrict each node $v_i$ to connect with at most $K$ most related nodes $\mathcal{N}_i$ that satisfy both conditions, where $K$ is a tunable parameter.
\begin{equation}
    \mathcal{N}_i=\textit{Top-K}_{j\neq i}(w_{i,j}).
\end{equation}
When introducing a new event node, entities extracted from it determine connections to existing nodes, naturally embedding the event into meaningful semantic and causal contexts. This incremental insertion enriches the event graph dynamically, enabling coherent growth and the continuous extension of event chains, effectively capturing the evolving dynamics intrinsic to the DyG.

\textbf{(3) Event Graph Indexing.}
To balance semantic relevance and temporal proximity, we extend \textsc{NanoVectorDB}\footnote{\url{https://github.com/gusye1234/nano-vectordb}} with a \texttt{TimestampEnhancedVectorStorage} layer that concatenates a sinusoidally encoded timestamp $\mathbf{h}_i^{time}\!\in\!\mathbb{R}^{d_\tau}$ to each DEU semantic embedding $\mathbf{h}_i^{text}\!\in\!\mathbb{R}^{d}$, producing a time-enhanced embedding $\mathbf{z}_i\!\in\!\mathbb{R}^{d+d_\tau}$ that is stored in the vector index. The event graph $G=(\mathcal{V},\mathcal{E})$ is managed using \textsc{NetworkX}\footnote{\url{https://networkx.org/}}, where edge weights $w_{i,j}$ are preserved for traversal and a bidirectional ID map links vector search results to graph nodes.

By combining semantic and temporal signals, the event graph reflects narrative flow with meaningful path semantics: each hop represents a grounded, interpretable transition between events. This design also supports parallel, multi-path traversal strategies in retrieval, enabling \method~ to reconstruct temporally consistent and semantically coherent event chains for complex queries (as introduced in Section~\ref{sec:retrieval}).

\subsection{Event Timeline Retrieval and Prompting with Time CoT}
\label{sec:retrieval}

\textbf{(1) Query Parse with Temporal Intent.}
To enable time-aware retrieval over the event graph, we design a query processing module that transforms natural language questions into joint semantic-temporal embeddings. This allows queries to align not only with topical content but also with temporal intent.

Given a user query $Q$, we perform two steps: temporal information extraction and query embedding. 
For the temporal information extraction, we leverage an LLM to extract temporal constraints $t_{Q}$ from the query text. 
Then we encode the query by separately extracting its semantic and temporal representations:
\begin{equation}
 \mathbf{h}_{Q}^{text}={\rm Encoder}_{text}(Q),~\mathbf{h}_{Q}^{time}=\phi(t_Q),
\end{equation}
where $\textbf{h}_{Q}^{text}$ is the semantic embedding produced by the same encoder used for DEUs and $\textbf{h}_{Q}^{time}$ is the temporal encoding generated using the same Fourier-based mapping $\phi(\cdot)$ as in the event graph.
Before retrieval, we reweight the temporal component of the query embedding by a tunable factor $\lambda$, producing:
\begin{equation}
\tilde{\mathbf{h}}_Q(\lambda) = \bigl[\mathbf{h}_Q^{\text{text}};\,\lambda\,\mathbf{h}_Q^{\text{time}}\bigr],
sim(Q,i) = \cos\bigl(\tilde{\mathbf{h}}_Q(\lambda),\,\mathbf{h}_i\bigr),
\end{equation}
Where \(\lambda \in [0,1]\) controls the trade-off between semantic and temporal components, and \(i\in V\) indexes each candidate DEU node with embedding \(\mathbf{h}_i\).

By dynamically tuning the temporal scaling factor, \method~ retriever strikes an optimal balance between semantic similarity and temporal proximity, substantially boosting retrieval precision for time-sensitive queries. It also guards against selecting events that merely coincide in time but bear no semantic relation to the query.

\textbf{(2) Graph Traversal and Timeline Construction.}
Then the resulting vector $\tilde{\mathbf{h}}_Q$ is used to retrieve temporally aligned and semantically relevant event nodes from the vector database, forming the entry point for subsequent graph traversal and answer generation. To refine this initial set, a cross-encoder reranker scores each retrieved event against the query to filter out irrelevant or weakly aligned nodes, ensuring higher-quality seeds. \method~then selects a set of top-ranked nodes $\mathcal{V}_{seed}$ to serve as the initial event nodes for graph traversal. Given the seed node set $\mathcal{V}_{seed}$ retrieved from the vector database, \method~performs graph traversal to collect supporting evidence in the form of a coherent event sequence. 
A weighted random walk mechanism is used to explore the event graph while respecting semantic and temporal locality.
For each seed node $v_i\in \mathcal{V}_{seed}$, we initiate multiple random walks with fixed length $L$. 
At each step, the next node $v_{l+1}$ is sampled from the neighbors of $v_l$ according to an edge-weighted transition probability:
\begin{equation}
P(v_{l+1} = u \mid v_l) = \frac{w(v_l, u)}{\sum_{u' \in \mathcal{N}(v_l)} w(v_l, u')},
\label{transition-prob}
\end{equation}
where $w(v_l, u) \in [0, 1]$ denotes the edge weight encoding semantic and temporal relevance between nodes $v_l$ and $u$, as defined in Equation~\eqref{edge-weight}.

This stochastic traversal allows the model to discover multi-hop paths that encode causal chains, narrative progression, or temporally adjacent developments beyond immediate neighbors. 
The random walk process produces a diverse set of paths $\{\mathcal{P}_1,\mathcal{P}_2,\cdots,\mathcal{P}_L\}$, each representing a candidate sequence of related events.

We then convert the path set into a structured timeline. 
Specifically, we separate static events from timestamped ones to ensure that temporally ambiguous information and unchanged facts are isolated from concrete temporal anchors. The timestamped events are then chronologically sorted to preserve the natural order of occurrence. Both static and temporal events are merged into a unified, coherent sequence, with each entry explicitly formatted as \textit{``Event \# [index] [timestamp]: sentence''}, making temporal cues transparent to the model.
This event timeline is passed to the LLM as structured context, enabling it to perform fine-grained temporal reasoning (\textit{e.g.}, identifying intervals, detecting overlaps, and inferring state persistence across time points) within a clear and interpretable reasoning chain.

\textbf{(3) Time Chain-of-Thought for Prompt.} 
In standard RAG workflows, the retrieved chunks are concatenated into a single prompt without any specific order. 
As a result, LLMs must implicitly infer and reorder the timeline, which can lead to errors such as temporal inversion or redundant references. 
To address these challenges, we introduce Time Chain-of-Thought (\textbf{Time-CoT}), a specialized chain-of-thought prompt that expilicitly encodes temporal relations (\textit{e.g.}, intervals, containment, and causal succession), guiding LLMs through temporal verification and reasoning before answer generation.

\begin{figure*}[!tp]
  \centering
  \includegraphics[width=1\linewidth]{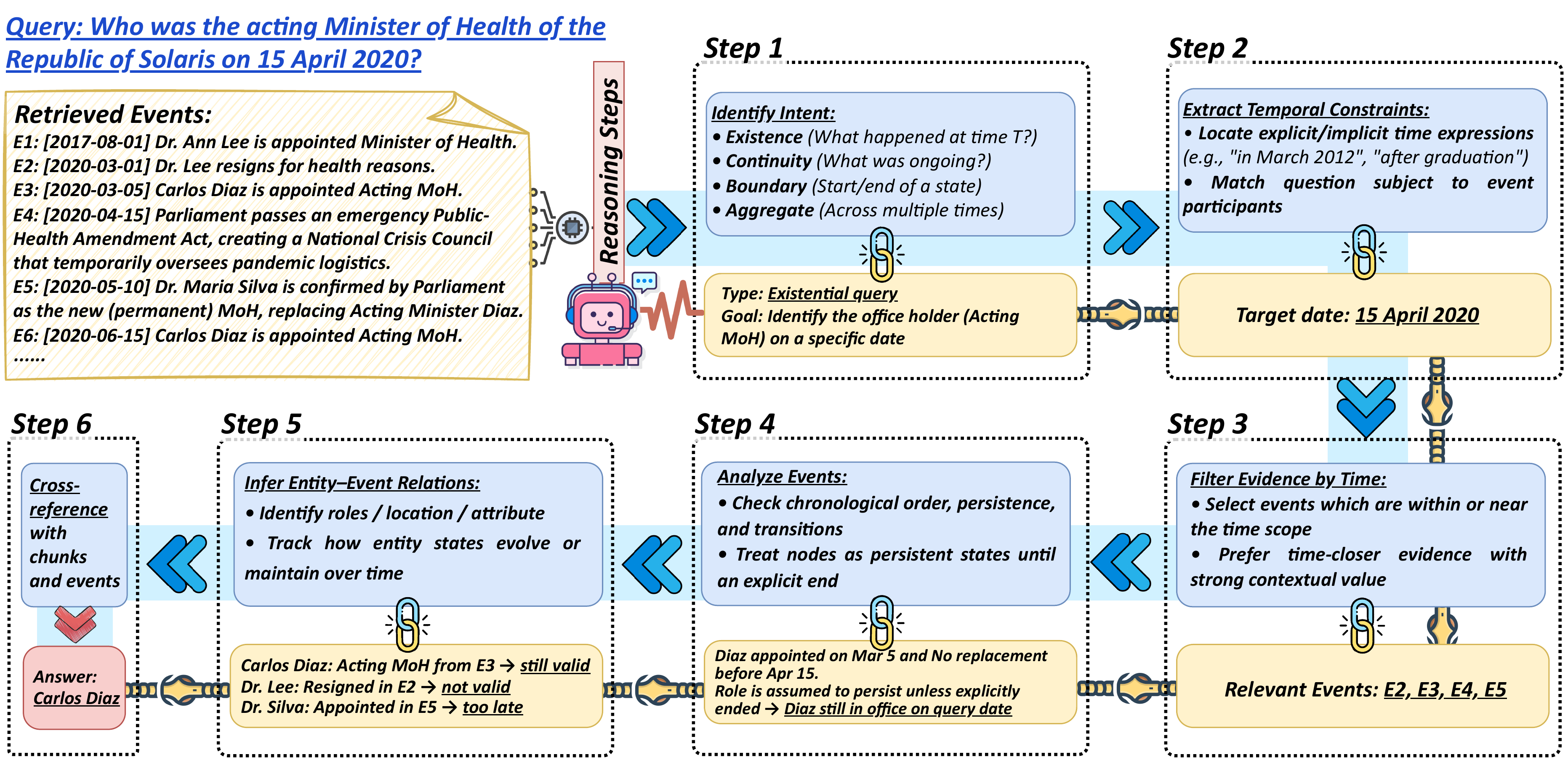}
  \caption{Illustration of Time-CoT. 
  }
\label{fig:CoT}
\end{figure*}

Chain-of-Thought prompting is known to boost multi-step reasoning by forcing the model to ``think'' step-by-step~\citep{CoT4LLM}, and our Time-CoT extends this benefit to temporal logic, explicitly guiding the LLM through interval comparison and persistence inference. Time-CoT enriches the LLM input with two key components: \textit{Structured Event Timeline} and \textit{Temporal Reasoning Template}. 
As illustrated in Figure~\ref{fig:CoT}, after obtaining the Structured Event Timeline, we apply a Temporal Reasoning Template to structure the prompt and explicitly guide the LLM through symbolic temporal inference. This template encodes procedural reasoning steps: The template first directs the model to identify evidence within the question’s time scope, using both explicit timestamps and relative expressions. It then encourages checking event order, tracking state continuity (\textit{e.g}, ``\textit{office incumbency}''), and distinguishing between event types such as instantaneous actions or ongoing processes. To further enhance precision, the template decomposes question semantics into predefined classes (\textit{e.g}, ``\textit{boundary}'', ``\textit{continuity}'', ``\textit{aggregate}''), and couples each with tailored reasoning heuristics. Finally, the prompt encourages cross-referencing between events and text chunks, enforcing justifications that cite specific timestamps and chains of events. This structured format converts flat retrieval results into a temporally aware reasoning process grounded in discrete, interpretable steps.

Time-CoT not only improves the model’s ability to handle complex temporal reasoning, mitigating temporal hallucination, but also yields a more interpretable reasoning process, as the chronological event chain makes the temporal logic transparent and grounded in the retrieved evidence.

\subsection{Discussion: DyG-RAG vs. TKG enhanced LLM Generation.}
Temporal Knowledge Graphs (TKGs) are often employed to inject temporal information into LLM generation by encoding facts as timestamped tuples  \textit{``$\langle$subject, relation, object, time$\rangle$''} as introduced in Section~\ref{sec:relatedwork}.
These symbolic structures are often integrated into generation pipelines via entity linking, temporal fact retrieval, or TKG embeddings. 
While effective in certain tasks such as complex QA and explainable temporal reasoning, TKG-enhanced generation faces several key limitations when compared with our dynamic graph retrieval-augmented generation framework,  \method.

\textbf{(1) Knowledge Granularity and Expressiveness.}
TKGs are fundamentally relation-centric and represent knowledge as atomic facts spanning time intervals. 
While effective for modeling persistent relations (\textit{e.g.}, \textit{``$\langle$Obama, presidentOf, US, 2009–2017$\rangle$''}), they lack expressiveness for capturing transient, state-changing events, causal chains, or multi-clause descriptions. 
In contrast, \method~introduces Dynamic Event Units (DEUs), defined as self-contained, time-anchored factual statements, offering much finer granularity and aligning naturally with how humans ask and reason about temporal questions.

\textbf{(2) Graph Construction and Context Adaptability.}
TKGs are typically constructed using a predefined schema, a fixed set of entity types and relation predicates curated from knowledge base ontologies. 
While this schema-driven approach ensures consistency, it also imposes significant limitations. 
The expressivity of TKGs is constrained by prior assumptions in the ontologies, making it difficult to capture emergent and domain-specific temporal information outside the schema. 
In contrast, \method~builds the event graph directly from free-form text, using data-driven DEUs without reliance on a pre-fixed schema. 
This allows the event graph to naturally adapt to the distribution of knowledge in the corpus and flexibly support context-dependent retrieval, enabling finer alignment with the LLM’s generation needs.

\textbf{(3) Grounding and Text Fidelity.} 
A critical limitation of many TKGs is that they are pre-constructed independently of the raw corpus, typically storing abstracted entity–relation triples without links to the original sentences. 
This detachment hinders provenance tracking, fine-grained context retrieval, and answer justification. 
In contrast, \method~preserves raw text grounding and context, which can be surfaced during generation to support transparency and trustworthiness.

In summary, while traditional TKGs offer structured storage of time-stamped facts, they struggle with expressiveness, adaptability, and grounding with text fidelity. 
In contrast, \method~builds an event graph from raw text using fine-grained dynamic event units, captures both semantic and temporal relations, and supports efficient multi-hop retrieval. 
This makes it more flexible, interpretable, and better suited for generating temporally grounded, context-aware answers.

\section{Experiments}
\label{sec:exp}
This section empirically evaluates the proposed \method~on the temporal question-answering (QA) task. 
The experiments focus on the following research questions: 
\begin{itemize}[leftmargin=*]
    \item \textbf{Q1.} 
    How does \method~perform in the temporal QA task? 
    (Section~\ref{sec:QA}) 
    \item \textbf{Q2.} 
    How critical is the event graph 
    for temporally grounded generation tasks? 
    (Section~\ref{sec:exp_eventgraph}) 
    \item \textbf{Q3.} 
    Can \method~effectively retrieve and reason over temporal event knowledge? 
    (Section~\ref{sec:exp_timeline}) 
    \item \textbf{Q4.} 
    How efficient is \method~in terms of index and query?
    (Section~\ref{sec:exp_efficiency})
\end{itemize}

\subsection{Experimental Settings}
\subsubsection{Datasets}
To evaluate \method’s ability to retrieve and reason over temporal information, we focus on three types of temporal questions: \textbf{Implicit Temporal Inference}, \textbf{Event State Grounding}, and \textbf{Multi-hop Temporal Reasoning}. 
Specifically, we conduct experiments on three open-source temporal QA benchmarks for the above three question types: TimeQA~\citep{chen2021timeqa}, TempReason~\citep{tan2023tempreaon}, and ComplexTR~\citep{tan2023complextr}. 
These datasets, derived from Wikipedia-based sources~\citep{wikipedia}, closely simulate temporal RAG settings by providing structured temporal annotations and factual content essential for constructing realistic question-answer pairs.
\begin{itemize}[leftmargin=1.5em]
\item \textbf{TimeQA}~\citep{chen2021timeqa}.
Focused on detailed reasoning with implicit temporal references and cross-sentence cues, we adopt the hard mode, where questions demand deeper inference rather than keyword matching.
\item \textbf{TempReason}~\citep{tan2023tempreaon}.
Centered on event state grounding, we use the L2 mode to evaluate the model’s retrieval ability in determining event states at specified timestamps.
\item \textbf{ComplexTR}~\citep{tan2023complextr}.
This benchmark targets multi-hop temporal reasoning, which requires combining multiple temporal cues across events. We use the golden test set to assess the model’s chaining capability.
\end{itemize}
\begin{table}[]
\caption{Statistics of Temporal QA Datasets. }
\label{tab:dataset}
\resizebox{\linewidth}{!}{
\centering
\renewcommand\arraystretch{1.8}
\begin{tabular}{lrrrrl}
\toprule
\textbf{Dataset} &
  \multicolumn{1}{r}{\textbf{Documents}} &
  \multicolumn{1}{r}{\textbf{Tokens}} &
  \multicolumn{1}{r}{\textbf{Avg. Tokens}} &
  \multicolumn{1}{r}{\textbf{Questions}} &
  \multicolumn{1}{l}{\textbf{Question Type}} \\ \midrule
TimeQA~\citep{chen2021timeqa}     & 3,159 & 1,819,999 & 576   & 2,613 & Implicit Temporal Inference  \\ \midrule
TempReason~\citep{tan2023tempreaon} & 2,721 & 2,667,820 & 980 & 5,397 & Event State Grounding  \\ \midrule
ComplexTR~\citep{tan2023complextr}  & 112   & 240,165   & 2,144 & 312   & Multi-hop Temporal Reasoning \\ \bottomrule
\end{tabular}
}
\end{table}
As no specialized open-source benchmark exists for temporal RAG, we adapt these datasets into a unified document corpus and structured question–answer pairs, reflecting real-world RAG pipelines for temporal retrieval and generation. We have made our processed datasets\footnote{\url{https://github.com/RingBDStack/DyG-RAG/datasets}} publicly available to facilitate further research.
The statistics of the dataset are shown in Table~\ref{tab:dataset}.

\subsubsection{Baselines}
To the best of our knowledge, there are no existing Dynamic Graph RAG works. 
For comparison, we select several RAG methods and Graph RAG methods. 
\textbf{RAG methods} include \textit{Vanilla RAG}.
\textbf{Graph RAG methods} include \textit{GraphRAG}~\citep{graphrag}, \textit{LightRAG}~\citep{lightrag}, \textit{E$^2$GraphRAG}~\cite{E2GraphRAG}, \textit{HippoRAG}~\citep{hipporag}.
\begin{itemize}[leftmargin=1.5em]
\item \textit{Vanilla RAG}\footnote{\url{https://huggingface.co/facebook/rag-token-nq}}~\citep{rag}: 
A standard baseline in simple RAG systems that
store the chunked texts in a vector database with text embeddings and use representation vectors to directly retrieve text chunks based on the similarity for the query.
\item \textit{GraphRAG}\footnote{\url{https://github.com/Graph-RAG/GraphRAG/}}~\citep{graphrag}: 
It transforms the retrieved nodes into communities and traverses the communities to capture the global information.
We use two variants of GraphRAG, including local and global retrieval, denoted as \textit{GraphRAG-L} and \textit{GraphRAG-G}.
\item \textit{LightRAG}\footnote{\url{https://github.com/HKUDS/LightRAG}}~\citep{lightrag}: It uses a dual-level retrieval mechanism to capture both low-level and high-level information of the index graph. We use three variants of LightRAG, including local, global, and hybrid retrieval, denoted as \textit{LightRAG-L}, \textit{LightRAG-G}, and \textit{LightRAG-H}.
\item \textit{E$^2$GraphRAG}\footnote{\url{https://github.com/YiboZhao624/E-2GraphRAG}}~\cite{E2GraphRAG}:
It constructs a summary tree and an entity graph based on document chunks and then constructs bidirectional indexes to capture their many-to-many relationships. 
\item \textit{HippoRAG}\footnote{\url{https://github.com/OSU-NLP-Group/HippoRAG}}~\citep{hipporag}:
It first constructs a knowledge graph for offline indexing and then uses personalized PageRank to identify relevant subgraphs for multi-hop reasoning.
\end{itemize}

\subsubsection{Settings}
To ensure consistency across baselines and fair comparison, we adopt the following unified settings for model architecture, document chunking, and retrieval parameters.
For the backbone models, we employ \texttt{Qwen2.5-14B}~\citep{qwen2} as the large language model for all graph-construction and generation tasks. And all retrieval modules use the \texttt{BGE-M3}~\citep{bgem3} encoder, a multilingual general-purpose embedding model, to encode both queries and text chunks into dense vector representations.

For our \method, entity extraction from document chunks is performed using the \texttt{dslim/bert-base-NER model}~\citep{bert}, which enables high-quality span-level tagging of person, organization, and location entities. And we use the lightweight but effective \texttt{TinyBERT-L-2-v2}~\citep{tinybert} as the initial retrieval candidate events reranker.
All documents are segmented using a sliding window approach with a fixed chunk size of 1,200 tokens and an overlap of 64 tokens. During retrieval, we select the Top-20 candidate chunks or nodes per query, and the input to the language model is truncated to a maximum of 16,384 tokens per instance.

We conduct all experiments using 4 NVIDIA V100 GPUs. Each GPU is equipped with 32 GB of HBM2 memory and delivers up to 15.7 TFLOPS of FP32 performance and 125 TFLOPS via Tensor Core FP16 operations. We deploy LLM inference using the vLLM framework to ensure efficient batching and low-latency throughput. We fix the maximum request concurrency to 32 in the graph construction stage if the method implementation supports concurrent LLM calls.

\subsection{Results on Temporal QA}
\label{sec:QA}
\begin{table}[!tp]
\caption{Results on temporal QA task, where Acc. means accuracy, \textcolor{crimson}{\textbf{Crimson}} denotes the best results and \textcolor{orange}{\textbf{Orange}} denotes the runner-ups.}
\label{tab:QAres}
\centering
\renewcommand\arraystretch{1.5}
\setlength{\tabcolsep}{2mm}{
\begin{tabular}{lrrrrrr}
\hline
\multirow{2}{*}{\textbf{Datasets}} & \multicolumn{2}{c}{\textbf{TimeQA}} & \multicolumn{2}{c}{\textbf{TempReaon}} & \multicolumn{2}{c}{\textbf{ComplexTR}} \\ \cline{2-7} 
                        & Acc. (\%) & Recall (\%) & Acc. (\%) & Recall (\%) & Acc. (\%) & Recall (\%) \\ \hline
Vanilla RAG             & 37.58     & 44.14       & 52.01     & 64.38       & 42.55     & 51.46       \\ \hline
GraphRAG-L              & 40.26     & 46.28       & 56.11     & 67.55       & 43.16     & 54.29       \\
GraphRAG-G              & 10.10     & 13.83       & 8.81     & 11.34        & 20.97     & 30.81       \\ \hline
LightRAG-L              & 34.94     & 39.89       & 54.23     & 66.82       & 41.03     & 49.46       \\
LightRAG-G              & 6.66      & 8.12        & 13.54     & 15.27       & 12.46     & 13.56       \\
LightRAG-H              & 36.36     & 43.06       & 50.01     & 62.95       & 42.68     & 53.59       \\ \hline
HippoRAG                & 39.99     & 45.39       & \textcolor{orange}{\textbf{69.80}}     & \textcolor{orange}{\textbf{80.54}}       & \textcolor{orange}{\textbf{44.68}}     & \textcolor{orange}{\textbf{55.28}}       \\ \hline
E$^2$GraphRAG              & \textcolor{orange}{\textbf{40.48}}     & \textcolor{orange}{\textbf{50.19}}       &   61.29    &   73.58   & 38.29     & 54.99       \\ \hline
\textbf{DyG-RAG (ours)} & \textcolor{crimson}{\textbf{58.78}}     & \textcolor{crimson}{\textbf{67.02}}       & \textcolor{crimson}{\textbf{84.75}}     & \textcolor{crimson}{\textbf{91.47}}       & \textcolor{crimson}{\textbf{55.62}}     & \textcolor{crimson}{\textbf{69.88}}       \\ \hline
\end{tabular}
}
\end{table}
To assess \method's effectiveness on the temporal-aware QA task, we follow the token-level evaluation protocol~\citep{indepth}, reporting Accuracy and Recall based on the overlap between predicted and gold-standard answer tokens.
The accuracy and recall on three benchmark datasets are shown in Table~\ref{tab:QAres}.
As we can see, \method~consistently achieves superior performance across all datasets, showing significant improvements in both accuracy and recall. Specifically, compared with the strongest baselines, \method~achieves absolute accuracy gains of approximately 18.30\%, 14.95\%, and 10.94\% on TimeQA, TempReason, and ComplexTR datasets, respectively. Correspondingly, recall improvements are about 16.83\%, 10.93\%, and 14.60\%.

The TimeQA dataset evaluates the model’s capability to handle complex temporal spans and hierarchical inclusion relations, which need reasoning in implicit temporal cues across sentences. 
\method~demonstrates superior performance primarily due to its dynamic event units (DEUs), which systematically organize temporal relationships through semantic-temporal links. 
This structured temporal arrangement significantly enhances the model’s capability to infer implicit temporal connections. 
And the auxiliary Time-CoT reasoning guides step-wise disambiguation of complex temporal connections. 
In contrast, baselines such as GraphRAG show unsatisfactory performance, as their static structures lack the granularity and adaptability required for effectively capturing temporal information.

The TempReason dataset assesses the accuracy of event state identification at specified timestamps. 
\method~excels in this task primarily due to its fine-grained DEU retrieval mechanism and precise time embeddings, enabling accurate retrieval of events aligned specifically to queried timestamps. 
By contrast, HippoRAG organizes information using concept nodes and employs PPR-guided multi-hop retrieval, which can diffuse relevance over loosely connected events and thus retrieve extraneous or time-mismatched states. 
Other methods that segment events at either overly fine-grained (entity-level) or overly coarse (chunk-level) granularity, which hinder the encoding of temporal discriminative information, omit essential context, or retrieve extraneous information, leading to retrieval errors and degraded event state grounding.

The ComplexTR dataset evaluates the capability for multi-hop temporal reasoning.
\method~achieves superior performance by seamlessly integrating DEUs into coherent semantic-temporal structures, effectively enabling robust multi-event reasoning across complex temporal contexts. 
Furthermore, the structured reasoning approach of Time-CoT supports coherent narrative chains over multiple temporal spans. 
Baseline methods such as LightRAG, which rely on keyword-based retrieval and shallow path aggregation, often miss important long-range event associations and lack dynamic re-weighting of emerging temporal links. 
Consequently, these approaches struggle to reliably construct robust multi-hop event sequences, resulting in inferior performance.

\subsection{Investigation on Event Graph Construction}
\label{sec:exp_eventgraph}

To better understand the role of event graph construction in \method, we conduct an ablation study focusing on the impact of different knowledge structuring paradigms. Specifically, we compare \method~with two alternative designs: \textbf{KG-based RAG}, which uses a static knowledge graph and temporal prompting, and \textbf{Chunk-based RAG}, which performs standard chunk-level retrieval without graph structure. We adopt different knowledge construction strategies and corresponding retrieval units, explicitly forcing the LLM to construct a timeline from the retrieved content before performing step-by-step reasoning to answer the question. 

\begin{figure}[!tp]
    \centering
    \includegraphics[width=1\linewidth]{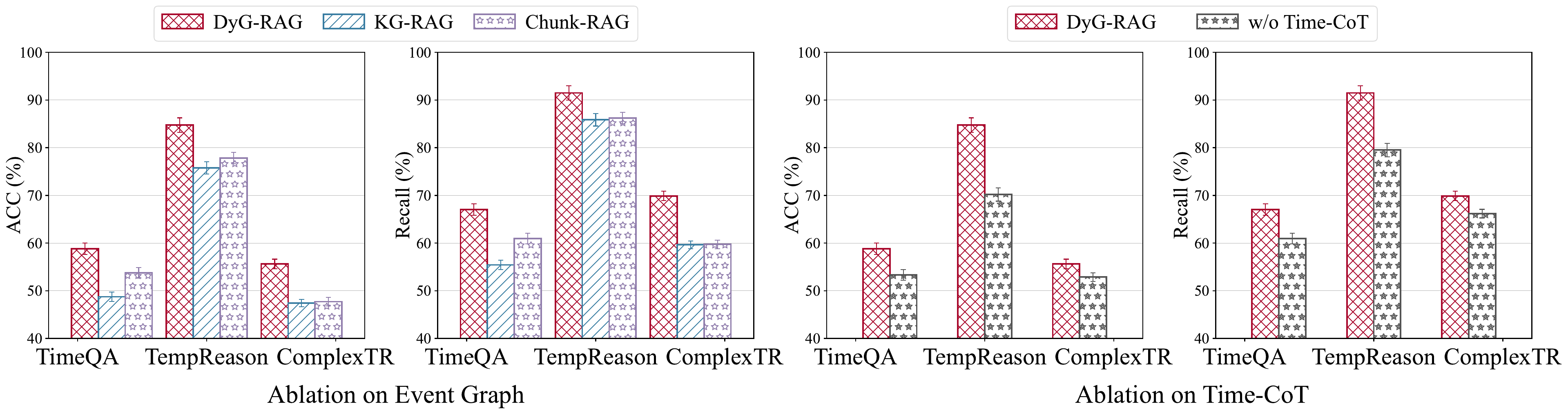}
    \caption{Ablation Study}
    \label{fig:ablation}
\end{figure}

As Figure \ref{fig:ablation}~shows, \method~consistently outperforms two knowledge construction methods across the three datasets, especially on ComplexTR, which requires multi-hop temporal reasoning. 
Unlike KG-RAG, which uses a static KG lacking explicit chronological relations, and Chunk-RAG, which retrieves isolated textual chunks without inherent temporal connectivity, our \method~inherently encodes events within a dynamic and temporally structured graph. 
This intrinsic event-level structuring enables the LLM to organize retrieved content into coherent and temporally consistent timelines, thereby substantially enhancing its capability to perform temporal reasoning.

\subsection{Investigation on Event Timeline and Time CoT}
\label{sec:exp_timeline}
To assess the contribution of temporal reasoning in \method, we conduct an ablation study in \method~ where the components of Time-CoT are removed. Specifically, we exclude the explicit event timeline construction from the retrieval outputs and omit the structured temporal reasoning instructions in the prompt.

As Figure \ref{fig:ablation}~shows, incorporating Event Timelines and Time-CoT prompting leads to consistently better performance across all datasets. This suggests that prompting the LLM to reason step by step over retrieved timeline events significantly enhances its ability to handle complex temporal questions. 
Instead of treating retrieved information as isolated facts, Time-CoT encourages LLM to construct a coherent chronological narrative and reason explicitly within this structure. Such guided reasoning aligns with the core principles of chain-of-thought prompting, helping LLM more effectively capture temporal dependencies and improve answer accuracy in time-sensitive tasks.


\subsection{Efficiency of \method}
\label{sec:exp_efficiency}

To assess the efficiency of~\method, we report the \textbf{Indexing Time (IT)} and \textbf{Querying Time (QT)} across three temporal QA datasets. We include a representative set of GraphRAG-based baselines, all of which adopt structured graph construction for retrieval-augmented generation. For consistency, indexing time is defined as the total time required for graph construction, while querying time refers to the average time per query, encompassing both retrieval from the constructed index and answer generation by LLM.

\begin{figure}[!tbp]
  \centering
  \includegraphics[width=\textwidth]{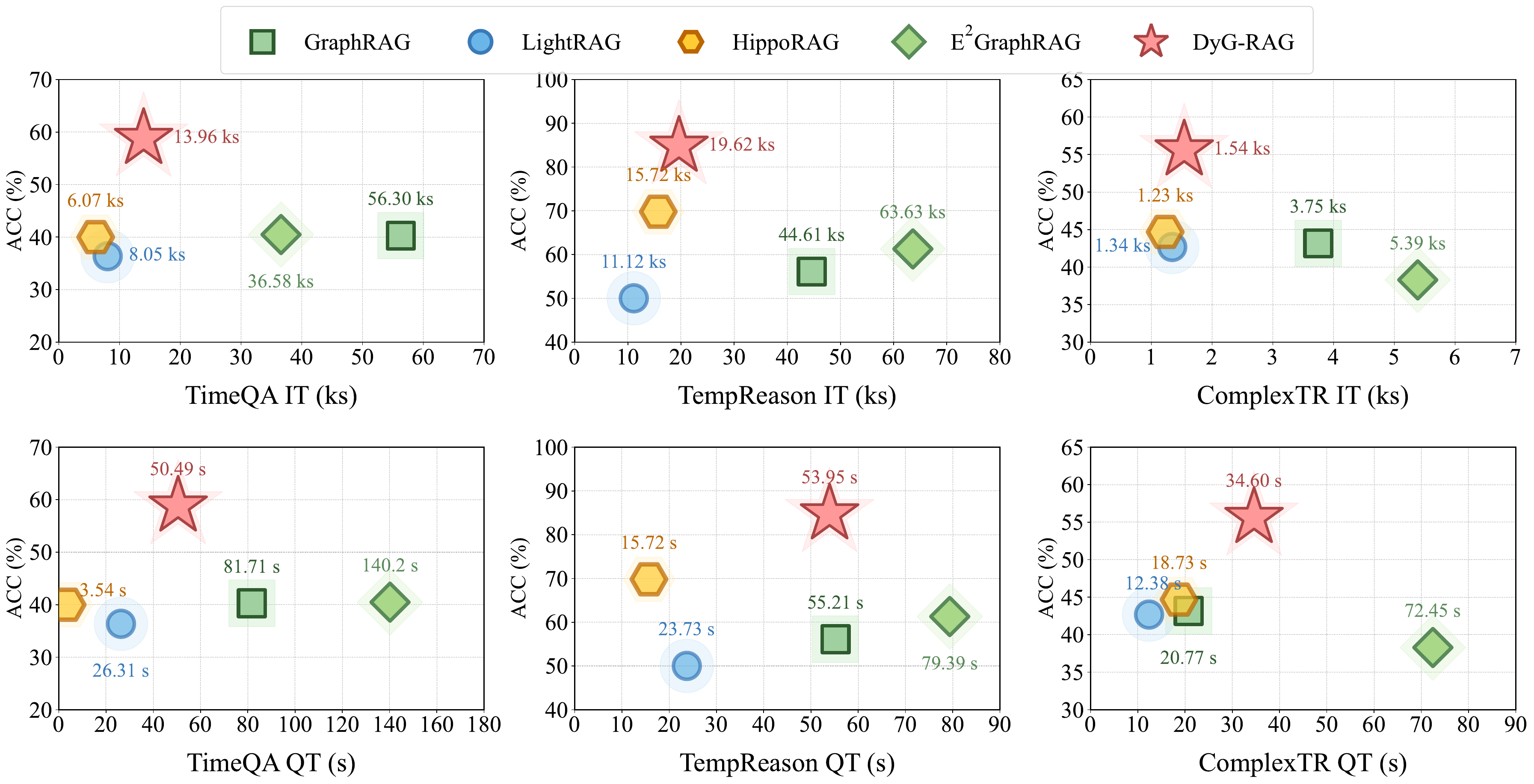} 
  \caption{Efficiency Comparison}
  \label{fig:efficiency}
\end{figure}

As Figure~\ref{fig:efficiency} shows, efficiency across evaluated methods varies significantly, reflecting differences in both algorithmic complexity and engineering maturity. HippoRAG and LightRAG achieve the fastest querying time and indexing time, primarily due to their mature, optimized code implementations and efficient pipeline design. In contrast, GraphRAG exhibits notably slower indexing performance, mainly attributed to the expensive community report generation phase, which involves intensive graph clustering and serialization steps. E$^2$GraphRAG introduces valuable acceleration strategies, particularly in entity extraction and retrieval; however, its end-to-end efficiency is hindered by the lack of concurrent LLM request implementation and lengthy answer-generation prompts. Our \method~demonstrates moderate yet acceptable latency, ranking consistently third-best overall. Considering its substantial accuracy improvements, the current efficiency trade-off is reasonable. Nonetheless, further speed optimization remains an important open issue for future work.

\subsection{Case Study}
To gain a deeper understanding of the underlying mechanisms behind retrieval-based temporal reasoning, we perform case studies of three representative methods:~\method, GraphRAG, and Naive RAG on the three Temporal QA datasets. These analyses illustrate how each method retrieves, structures, and reasons over temporal information, providing interpretable insights into their relative behaviors and failure modes.


\begin{figure}[htbp]
  \centering
  \includegraphics[width=\textwidth]{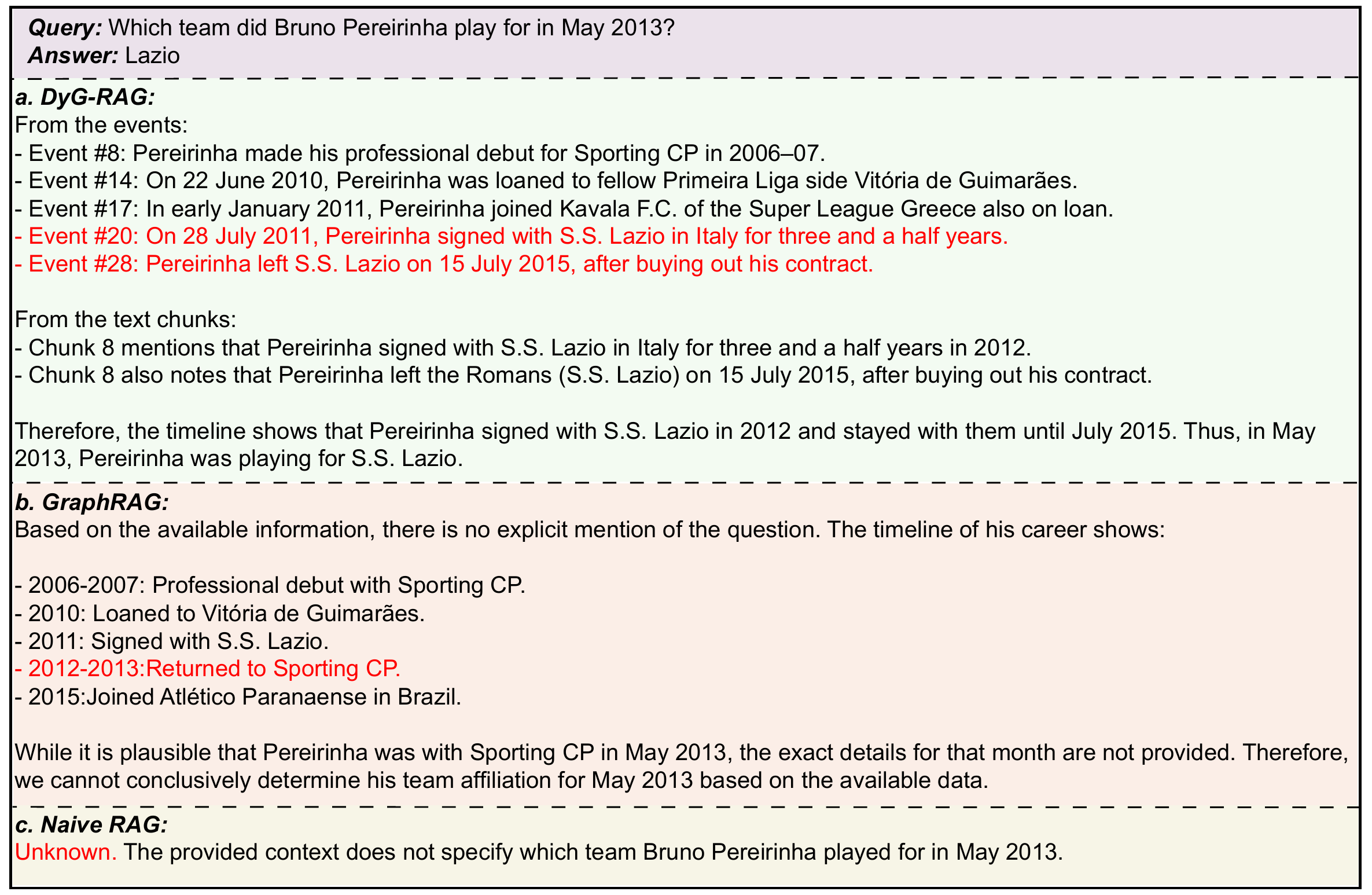}
  \caption{Case Study on TimeQA}
  \label{fig:case1}
  \vspace{-0.8em}
\end{figure}

Figure~\ref{fig:case1} illustrates the outputs of three representative methods on the TimeQA dataset, which involves tasks requiring implicit temporal reasoning. The specific query, \textit{``Which team did Bruno Pereirinha play for in May 2013?"}, exemplifies implicit temporal inference as correctly answering the question requires reasoning over two temporally distinct events rather than directly locating explicit mentions of the year 2013. \method~correctly identified \textit{``S.S. Lazio''} as the team due to its capability to build a coherent event chain, explicitly capturing both contract initiation and termination events and thus accurately inferring the implicit timeframe. In contrast, GraphRAG failed by hallucinating an incorrect event (\textit{``Returned to Sporting CP''}), caused by incomplete temporal attributes in its KG structure. Meanwhile, Naive RAG lacked the ability to retrieve or infer the necessary implicit temporal connections, leaving the query unanswered. This case illustrates \method's effectiveness in reliably modeling implicit temporal reasoning through dynamic event graph construction.

\begin{figure}[htbp]
  \centering
  \includegraphics[width=\textwidth]{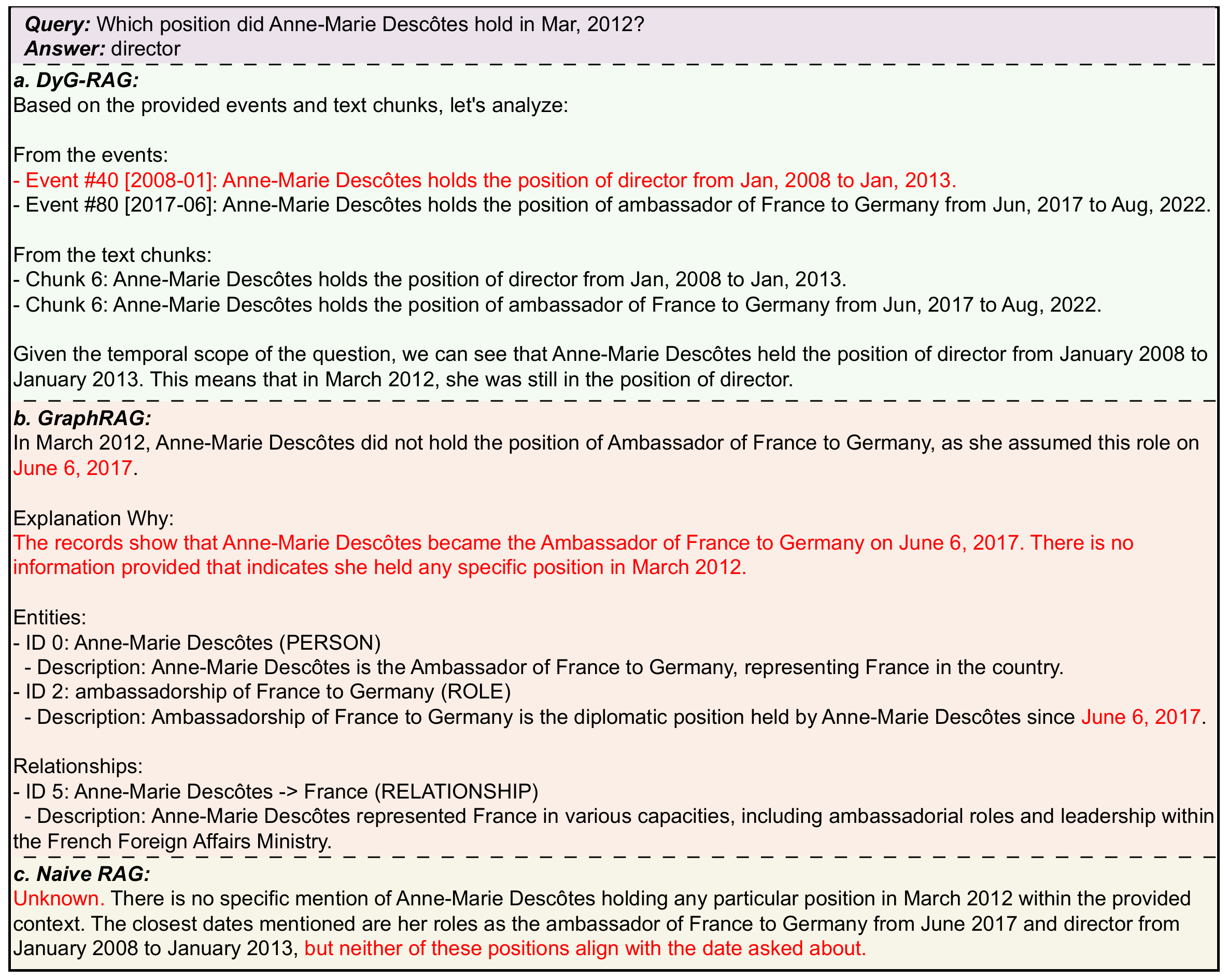}
  \caption{Case Study on TempReason}
  \label{fig:case2}
  \vspace{-0.8em}
\end{figure}

Figure~\ref{fig:case2} illustrates the outputs of three representative methods on the TempReason dataset. This example specifically demonstrates event state grounding, as the query \textit{``Which position did Anne-Marie Descôtes hold in Mar 2012?"} requires determining a person's exact professional state at a given timestamp, achievable by pinpointing a single relevant event interval. In this case, \method~successfully answers \textit{"director"} by leveraging its explicit time embeddings to accurately retrieve the event interval \textit{(2008--2013)} closely aligned with the queried timestamp \textit{(March 2012)}. \method~benefits further from structured event triples paired with textual chunks, allowing LLM to validate and reinforce its reasoning effectively. In contrast, GraphRAG incorrectly selects the later-held position (\textit{``Ambassador of France to Germany"}), primarily due to overly dense connectivity within its predefined KG around \textit{Anne-Marie Descôtes}. Its KG-based graph is constructed with nodes representing entities and positions, and edges encoding their relations, but without explicit temporal scopes or directional indicators of state progression. As a result, it cannot capture whether a state is being maintained or has changed over time, leading to confusion between temporally distant roles and misalignment with the queried time. Meanwhile, Naive RAG, despite correctly retrieving a text chunk containing relevant temporal information (\textit{``director from 2008 to 2013"}), fails because isolated and unstructured passages hinder LLM's ability to reliably infer persistence of states over time. Consequently, Naive RAG's lack of explicit temporal structuring causes LLM to discard the correct evidence, demonstrating the necessity of structured temporal grounding as provided by \method. 

\begin{figure}[htbp]
  \centering
  \includegraphics[width=\textwidth]{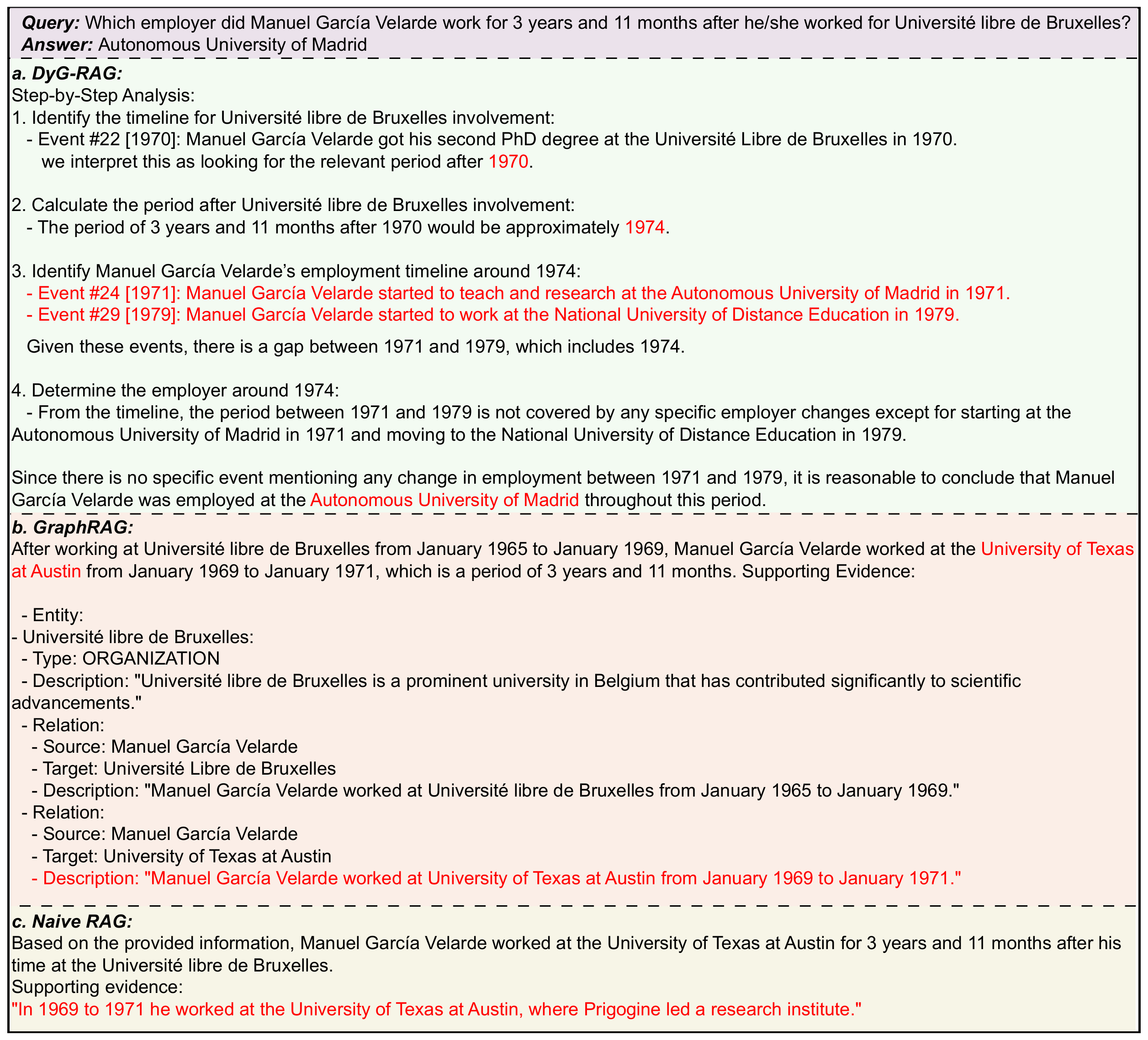}
  \caption{Case Study on CompexTR}
  \vspace{-0.8em}
  \label{fig:case3}
\end{figure}

Figure~\ref{fig:case3} illustrates the outputs of three representative methods on the ComplexTR dataset. This query exemplifies multi-hop temporal reasoning, as it requires bridging two semantically distinct event types: an academic milestone (\textit{``pursuing a second doctorate at \underline{Université libre de Bruxelles} in 1970''}) and a subsequent employment event occurring 3 years and 11 months later. Among the compared methods, only \method~correctly identified (\textit{``Autonomous University of Madrid} '') as the answer, benefiting from its temporal-semantic graph structure that naturally supports reasoning across temporally distant events. By explicitly encoding both temporal and semantic relations, \method~enables the construction of a coherent event timeline, allowing the LLM to perform step-by-step inference over long-range dependencies. In contrast, both GraphRAG and Naive RAG failed for a common reason: they retrieved the nearest temporal event \textit{(``1969--1971, University of Texas at Austin''}) and prematurely terminated the reasoning process, exhibiting early stopping behavior. GraphRAG, despite its multi-hop capability, lacks temporal awareness in its edge construction, limiting traversal to semantically close nodes while ignoring chronological progression. As a result, both baselines missed critical events spanning a broader temporal window, hindering the LLM's ability to resolve the query through temporal chaining. 
\section{Conclusion}
\label{sec:conclusion}
In this work, we propose \method, a dynamic graph retrieval-augmented generation framework that addresses the limitations of existing RAG methods in temporal reasoning. 
By transforming unstructured text into a dynamic event graph, DyG-RAG captures the evolving structure of knowledge and supports precise, time-aware retrieval. 
It identifies semantically and temporally grounded event nodes through event unit extraction, encodes entity and temporal relations to model narrative and causal flows via graph indexing, and reconstructs coherent event sequences through structured retrieval. 
By enabling fine-grained temporal understanding and event-level reasoning, \method~lays a foundation for more faithful, context-aware language models in dynamic, real-world domains.

\medskip
\bibliographystyle{unsrt}
\bibliography{ref}

\end{document}